\documentclass[12pt,notitlepage]{article}
\usepackage{bbm}
\usepackage{dsfont}
\usepackage[T1]{fontenc}
\usepackage{lmodern}
\usepackage[active]{srcltx}
\usepackage{amsmath,amssymb,amsfonts}
\usepackage{hyperref}
\usepackage[english]{babel}
\usepackage[latin1]{inputenc}

\title{The Jones vector as a spinor and its representation on the Poincar\'e
sphere} 
\author{{\small \bf G.F. Torres del Castillo$^{1}$ and I. Rubalcava-Garc\'ia$^{2}$ \footnote{Present address: Instituto de F\'isica y Matem\'aticas, Universidad Michoacana de San Nicol\'as de Hidalgo, Ciudad Universitaria, Morelia, Mich., 58040, M\'exico and Centro de Ciencias Matem\'aticas UNAM, A.P. 61-3, Morelia, Mich., 58090, M\'exico.
Email: irais@ifm.umich.mx, irais@matmor.unam.mx}  } \\
{\small \it $^{1}$Departamento de F\'{\i}sica Matem\'atica, Instituto de Ciencias} \\
{\small \it Universidad Aut\'onoma de Puebla, 72570 Puebla, Pue., M\'exico.}\\
{\small \it $^{2}$ Facultad de Ciencias F\'{\i}sico Matem\'aticas}\\
{\small \it Universidad Aut\'onoma
de Puebla, Apartado postal 1152, 72001 Puebla, Pue., M\'exico. }}
\date{}

\begin{document}


\maketitle

\begin{abstract}
It is shown that the two complex Cartesian components of the electric field of a
monochromatic electromagnetic plane wave, with a temporal and spatial dependence of the
form ${\rm e}^{{\rm i} (kz - \omega t)}$, form a SU(2) spinor that corresponds to a
tangent vector to the Poincar\'e sphere representing the state of polarization and
phase of the wave. The geometrical representation on the Poincar\'e sphere of the
effect of some optical filters is reviewed. It is also shown that in the case of a
partially polarized beam, the coherency matrix defines two diametrically opposite
points of the Poincar\'e sphere.
\end{abstract}


{\bf Pacs:} 02.20.Qs, 03.50.De, 42.25.Ja\\


\section{Introduction}
In a recent paper \cite{GF} it has been shown that the (real) Cartesian components of
the electric field of a monochromatic electromagnetic plane wave can be expressed in
terms of a two-component SU(2) spinor, which specifies the amplitude, state of
polarization, and phase of the wave in such a way that two real mutually orthogonal
vectors made out of this spinor define the point of the Poincar\'e sphere corresponding
to the state of polarization and a tangent vector to the Poincar\'e sphere that
determines the phase of the wave. Furthermore, the inner product of the spinors
corresponding to two of these waves with the same wavevector (which is related to the
parallel transport of tangent vectors to the Poincar\'e sphere along a great circle arc
\cite{WP,GF}), determines if the waves are in phase according to Pancharatnam's
definition \cite{Pa}. (The relationship between the inner product of spinors and the
parallel transport along geodesics of the sphere was already recognized in Payne's 1952
paper \cite{WP}, without developing, however, its relationship with the interference of
electromagnetic waves. See also Ref.\ 4.)

The fact that the amplitude, state of polarization, and phase of a monochromatic
electromagnetic plane wave can be represented by a two-component spinor allows us to
derive many useful relations employing the same formalism as in Quantum Mechanics
\cite{GF}, instead of the not so widely known results of spherical trigonometry
\cite{Pa} (see also Ref.\ 5).

The state of polarization of a wave is usually specified making use of the Stokes
parameters or the Jones vector (see, {\it e.g.}, Refs.\ 6--11). The Stokes parameters
can be expressed in terms of the two-component spinor mentioned above \cite{GF} and, as
we shall show below, the Jones vector is essentially this spinor, expressed in an
appropriate basis.

In Sec.\ 2 we give a summary of the relevant results of Ref.\ 1, relating them with the
definition of the Jones vector. We show that, apart from the phase factor that gives
the time and space dependence of the electric field, the Jones vector is a
two-component spinor on which the rotations on the Poincar\'e sphere act through the
spin-1/2 representation. In Sec.\ 3 we review the effect of some optical filters and
its geometrical representation on the Poincar\'e sphere. We show that the effect of a
phase shifter corresponds to a rotation of the Poincar\'e sphere, while that of an
attenuator corresponds to a conformal transformation of this sphere (see also Refs.\ 10
and 11). In Sec.\ 4 we consider partially polarized beams, showing that the Stokes
parameters can be arranged into a $2 \times 2$ matrix that, except in the case of
unpolarized light, defines two diametrically opposite points of the Poincar\'e sphere.

Although some of the results obtained in this paper, such as the matrix form for phase
shifters and attenuators, are found in the literature using other approaches (see, {\em
e.g.}, Ref.\ 12 and the references cited therein), one remarkable feature of the spinor
formalism is that, besides the state of polarization represented by a point of the
Poincar\'e sphere, we also have the phase of the wave through the direction of a
tangent vector to the sphere at that point, which is not included in other approaches.
Thus the action of the optical filters are transformations not only on the points of
the Poincar\'e sphere, but also on the tangent vectors to this sphere.

\section{The Poincar\'e sphere}
The Cartesian components of the electric field of a monochromatic electromagnetic plane
wave propagating in the $z$-direction in a dielectric medium are usually expressed in
the form
\begin{equation}
E_{x} = {\rm Re} \big\{ A_{1} \exp [{\rm i} (kz - \omega t + \phi_{1})] \big\}, \qquad
E_{y} = {\rm Re} \big\{ A_{2} \exp [{\rm i} (kz - \omega t + \phi_{2})] \big\},
\label{usual}
\end{equation}
where $A_{1}$, $A_{2}$ are real, positive constants, $\omega$ and $k$ are the angular
frequency and wave number of the wave, respectively. At each point of space, the
resulting electric field describes an ellipse centered at the origin and, therefore,
the (real) electric field can be conveniently written as
\begin{eqnarray}
{\bf E} & = & \left[ a \cos {\textstyle \frac{1}{2}} \phi \, \cos (\omega t - kz +
{\textstyle \frac{1}{2}} \chi) - b \sin {\textstyle \frac{1}{2}} \phi \, \sin (\omega t
- kz + {\textstyle \frac{1}{2}}
\chi) \right] \hat{x} \nonumber \\
& & \mbox{} + \left[ a \sin {\textstyle \frac{1}{2}} \phi \, \cos (\omega t - kz +
{\textstyle \frac{1}{2}} \chi) + b \cos {\textstyle \frac{1}{2}} \phi \, \sin (\omega t
- kz + {\textstyle \frac{1}{2}} \chi) \right] \hat {y}, \label{2}
\end{eqnarray}
where $a$, $b$ are real constants, with $|a| \geqslant |b|$, $|a|$ is the major
semiaxis of the ellipse, $|b|$ is the minor semiaxis, and $\phi/2$ is the angle made by
the major axis of the ellipse with the $x$-axis, so that it suffices to consider values
of $\phi$ between 0 and $2 \pi$. The phase $\chi/2$ is necessary when one considers the
superposition of two or more waves \cite{GF}.

Since $|b/a| \leqslant 1$, for each value of the ellipticity, $b/a$, there is a unique
$\theta \in [0, \pi]$ such that
\[
\frac{b}{a} = \tan \left( \frac{\pi}{4} - \frac{\theta}{2} \right).
\]
Hence,
\begin{equation}
a = \sqrt{2} A \cos \left( \frac{\pi}{4} - \frac{\theta}{2} \right) = A (\cos
{\textstyle \frac{1}{2}} \theta + \sin {\textstyle \frac{1}{2}} \theta), \quad b =
\sqrt{2} A \sin \left( \frac{\pi}{4} - \frac{\theta}{2} \right) = A (\cos {\textstyle
\frac{1}{2}} \theta - \sin {\textstyle \frac{1}{2}} \theta), \label{3}
\end{equation}
for some constant $A$, which, with no loss of generality, we can assume positive. In
this way, $a \geqslant 0$, while $b$ is positive for $0 \leqslant \theta < \pi/2$ (in
which case the wave has right-hand polarization) and $b$ is negative for $\pi/2 <
\theta \leqslant \pi$ (then the wave has left-hand polarization). The values $\theta =
0$ and $\theta = \pi$ correspond to circular polarization, while $\theta = \pi/2$ in
the case of linear polarization. Making use of Eq.\ (\ref{3}), Eq.\ (\ref{2}) can be
rewritten in the form
\begin{eqnarray}
{\bf E} & = & A \left\{ \left[ \cos {\textstyle \frac{1}{2}} \theta \, \cos (\omega t -
kz + {\textstyle \frac{1}{2}} \chi + {\textstyle \frac{1}{2}} \phi) + \sin {\textstyle
\frac{1}{2}} \theta \, \cos (\omega t - kz + {\textstyle \frac{1}{2}} \chi -
{\textstyle \frac{1}{2}} \phi) \right] \hat{x} \right. \nonumber \\
& & \mbox{} + \left. \left[ \cos {\textstyle \frac{1}{2}} \theta \, \sin (\omega t - kz
+ {\textstyle \frac{1}{2}} \chi + {\textstyle \frac{1}{2}} \phi) - \sin {\textstyle
\frac{1}{2}} \theta \, \sin (\omega t - kz + {\textstyle \frac{1}{2}} \chi -
{\textstyle \frac{1}{2}} \phi) \right] \hat{y} \right\}. \label{4}
\end{eqnarray}

The parametrization of the electric field given by Eq.\ (\ref{4}) contains the same
number of independent parameters as expressions (\ref{usual}) (four real parameters).
However, by contrast with (\ref{usual}), the parameters appearing in Eq.\ (\ref{4})
specify more directly the polarization state of the wave [via Eqs.\ (\ref{3})].
Furthermore, by considering the angles $\theta$ and $\phi$ as spherical coordinates in
the usual manner ({\it i.e.}, $\theta$ as the polar angle and $\phi$ as the azimuthal
angle), each pair of values $(\theta, \phi)$ defines a point of the Poincar\'e sphere
\cite{BW,Gu,KL}.

Another set of parameters commonly employed to specify the polarization of a wave is
given by the Stokes parameters, $s_{0}$, $s_{1}$, $s_{2}$, $s_{3}$, which are related
to the angles $\theta$ and $\phi$ by means of \cite{BW,Gu} (see also Ref.\ 3 and the
references cited therein)
\begin{equation}
(s_{1}, s_{2}, s_{3}) = s_{0} (\sin \theta \cos \phi, \sin \theta \sin \phi, \cos
\theta), \label{stokes}
\end{equation}
where $s_{0}$ is the total flux density. Hence, $(s_{1}, s_{2}, s_{3})/s_{0}$ is the
point of the Poincar\'e sphere that corresponds to the polarization of the wave.

In the Jones formalism, the {\em complex}\/ Cartesian components of the electric field
form a column matrix (see, {\it e.g.}, Ref.\ 11 and the references cited therein),
\begin{equation}
\left( \begin{array}{c} E^{\rm c}_{x} \\ E^{\rm c}_{y} \end{array} \right) = \left(
\begin{array}{c} A \exp [{\rm i} (kz - \omega t + \phi_{1})] \\ B \exp [{\rm i} (kz -
\omega t + \phi_{2})]
\end{array} \right), \label{jones}
\end{equation}
where $A$ and $B$ are real constants. (We employ the superscript c in the components of
the electric field to emphasize the fact that they are complex.)

\subsection{Two-component spinors}
From Eq.\ (\ref{4}) we see that the components of the electric field are given by the
compact expression
\begin{equation}
E_{x} + {\rm i} E_{y} = A \left[ \cos {\textstyle \frac{1}{2}} \theta \, {\rm e}^{{\rm
i} (\omega t - kz + \chi/2 + \phi/2)} + \sin {\textstyle \frac{1}{2}} \theta \, {\rm
e}^{- {\rm i} (\omega t - kz + \chi/2 - \phi/2)} \right], \label{elec}
\end{equation}
or, in terms of the unit two-component spinor
\begin{equation}
o = \left( \begin{array}{c} o^{1} \\ o^{2} \end{array} \right) = {\rm e}^{- {\rm i}
\chi/2} \left( \begin{array}{c} {\rm e}^{- {\rm i} \phi/2} \cos \frac{1}{2} \theta \\
{\rm e}^{{\rm i} \phi/2} \sin \frac{1}{2} \theta
\end{array} \right), \label{7}
\end{equation}
we have
\begin{equation}
E_{x} + {\rm i} E_{y} = A ( {\rm e}^{{\rm i} (\omega t - kz)} \overline{o^{1}} + {\rm
e}^{- {\rm i} (\omega t - kz)} o^{2}), \label{9}
\end{equation}
where the bar denotes complex conjugation.

The two-component spinor (\ref{7}) may be familiar from Quantum Mechanics; it is the
normalized eigenspinor with eigenvalue $+ \hbar/2$ of the spin projection along the
direction with angles $\theta$, $\phi$. The unit spinor $o$ defines two mutually
orthogonal vectors with Cartesian components
\begin{equation}
R_{i} \equiv o^{\dag} \sigma_{i} o, \qquad M_{i} \equiv o^{\rm t} \varepsilon
\sigma_{i} o, \label{vectors}
\end{equation}
where $o^{\dag}$ is the transpose conjugate of $o$, $o^{\rm t}$ denotes the transpose
of $o$, the $\sigma_{i}$ are the standard Pauli matrices, and $\varepsilon \equiv
\left( \begin{array}{rc} 0 & 1 \\ -1 & 0 \end{array} \right)$ \cite{WP,3D}. The vector
$R_{i}$ is real and is the point of the Poincar\'e sphere that represents the
polarization state of the wave, {\it i.e.}, $(R_{1}, R_{2}, R_{3}) = (\sin \theta \cos
\phi, \sin \theta \sin \phi, \cos \theta)$. Hence, the Stokes parameters are directly
related to the unit spinor $o$ by
\begin{equation}
\frac{s_{i}}{s_{0}} = o^{\dag} \sigma_{i} o. \label{11}
\end{equation}
The direction of ${\rm Re}\, M_{i}$ does depend on the phase $\chi$ and, therefore,
$R_{i}$ together with ${\rm Re}\, M_{i}$ represent the state of polarization and the
phase of the wave \cite{GF}. Since ${\rm Re}\, M_{i}$ is orthogonal to $R_{i}$, ${\rm
Re}\, M_{i}$ is a tangent vector to the Poincar\'e sphere (${\rm Re}\, M_{i}$ forms an
angle $\chi$ with the meridian passing through the point $R_{i}$). In this manner, the
vector $R_{i}$, gives the point of the Poincar\'e sphere corresponding to the
polarization state of the wave, and ${\rm Re}\, M_{i}$ can be viewed as a tangent
vector to the Poincar\'e sphere, whose direction gives the phase of the wave.

If $o' = Qo$, with $Q \in {\rm SU(2)}$, then $o'$ is also a unit spinor and the vectors
$R'_{i}$ and $M'_{i}$, defined by $o'$, are related to $R_{i}$ and $M_{i}$,
respectively, by means of the SO(3) transformation, $(a_{ij})$, given by $Q^{\dag}
\sigma_{i} Q = \sum_{j = 1}^{3} a_{ij} \sigma_{j}$; that is $R'_{i} = \sum_{j = 1}^{3}
a_{ij} R_{j}$, and $M'_{i} = \sum_{j = 1}^{3} a_{ij} M_{j}$. Hence, each $Q \in {\rm
SU(2)}$ gives rise to a rotation on the Poincar\'e sphere. Conversely, given a rotation
on the Poincar\'e sphere, there exists a $Q \in {\rm SU(2)}$, defined up to sign,
corresponding to the rotation.

\subsection{Two spinor bases}
Since the components $E_{x}$ and $E_{y}$ appearing in Eq.\ (\ref{9}) are real, Eq.\
(\ref{9}) is equivalent to
\begin{equation}
E_{x} - {\rm i} E_{y} = A \big( {\rm e}^{{\rm i} (\omega t - kz)} \overline{o^{2}} +
{\rm e}^{- {\rm i} (\omega t - kz)} o^{1} \big). \label{9c}
\end{equation}
Hence, from Eqs.\ (\ref{9}) and (\ref{9c}) we see that
\begin{equation}
E_{x} = {\rm Re}\, \{ A [{\rm e}^{{\rm i} (kz - \omega t)} (o^{1} + o^{2})] \}, \quad
E_{y} = {\rm Re}\, \{ A [{\rm e}^{{\rm i} (kz - \omega t)} ({\rm i} o^{1} - {\rm i}
o^{2})] \} \label{complex}
\end{equation}
and, therefore, the components of the electric field are the real part of the {\em
complex}\/ functions $E^{\rm c}_{x}$, $E^{\rm c}_{y}$, given by the Jones vector
\begin{equation}
\left( \begin{array}{c} E^{\rm c}_{x} \\ E^{\rm c}_{y} \end{array} \right) = \sqrt{2}
\, {\rm e}^{- {\rm i} \pi/4} A {\rm e}^{{\rm i} (kz - \omega t)} \frac{{\rm e}^{{\rm i}
\pi/4}}{\sqrt{2}} \left(
\begin{array}{cr} 1 & 1 \\ {\rm i} & - {\rm i} \end{array} \right) \left(
\begin{array}{c} o^{1}
\\ o^{2} \end{array} \right) \label{rel}
\end{equation}
[{\it cf.}\ Eq.\ (\ref{jones})].

One can readily verify that the $2 \times 2$ matrix
\begin{equation}
\mathcal{U} \equiv \frac{{\rm e}^{{\rm i} \pi/4}}{\sqrt{2}} \left(
\begin{array}{cr} 1 & 1 \\ {\rm i} & - {\rm i} \end{array} \right) =
\frac{1}{2} \left( \begin{array}{cr} 1 + {\rm i} & 1 + {\rm i} \\ -1 + {\rm i} & 1 -
{\rm i} \end{array} \right), \label{u}
\end{equation}
appearing in Eq.\ (\ref{rel}), belongs to SU(2) and that
\begin{equation}
\mathcal{U} \sigma_{1} \mathcal{U}^{-1} = \sigma_{3}, \qquad \mathcal{U} \sigma_{2}
\mathcal{U}^{-1} = \sigma_{1}, \qquad \mathcal{U} \sigma_{3} \mathcal{U}^{-1} =
\sigma_{2}. \label{paulim}
\end{equation}
This means that $\mathcal{U}$ corresponds to a SO(3) transformation that permutes the
coordinate axes, $X, Y, Z$, of the Poincar\'e sphere and that, apart from the factor
$\sqrt{2} \, {\rm e}^{- {\rm i} \pi/4} A {\rm e}^{{\rm i} (kz - \omega t)}$, the Jones
vector $\left(
\begin{array}{c} E^{\rm c}_{x} \\ E^{\rm c}_{y} \end{array} \right)$
is essentially the two-component spinor $\left( \begin{array}{c} o^{1} \\ o^{2}
\end{array} \right)$ in a different basis. That is, letting
\begin{equation}
\left( \begin{array}{c} \tilde{o}^{1} \\ \tilde{o}^{2} \end{array} \right) \equiv
\mathcal{U} \left( \begin{array}{c} o^{1} \\ o^{2}
\end{array} \right), \label{relation}
\end{equation}
from Eq.\ (\ref{rel}) we have
\begin{equation}
\left( \begin{array}{c} E^{\rm c}_{x} \\ E^{\rm c}_{y} \end{array} \right) = \sqrt{2}
\, {\rm e}^{- {\rm i} \pi/4} A {\rm e}^{{\rm i}
(kz - \omega t)} \left( \begin{array}{c} \tilde{o}^{1} \\
\tilde{o}^{2}
\end{array} \right). \label{relb}
\end{equation}

While the basis spinors
\begin{equation}
\left( \begin{array}{c} o^{1} \\ o^{2} \end{array} \right) = \left(
\begin{array}{c} 1 \\ 0 \end{array} \right) \quad {\rm and} \quad \left(
\begin{array}{c} o^{1} \\ o^{2} \end{array} \right) = \left(
\begin{array}{c} 0 \\ 1 \end{array} \right) \label{circb}
\end{equation}
(which correspond to $\theta = 0$ and $\theta = \pi$, respectively) represent
circularly polarized waves, the basis spinors
\begin{equation}
\left( \begin{array}{c} \tilde{o}^{1} \\ \tilde{o}^{2}
\end{array} \right) = \left( \begin{array}{c} 1 \\ 0 \end{array}
\right) \quad {\rm and} \quad \left( \begin{array}{c} \tilde{o}^{1} \\
\tilde{o}^{2} \end{array} \right) = \left( \begin{array}{c} 0 \\ 1
\end{array} \right) \label{linb}
\end{equation}
represent linearly polarized waves [see Eq.\ (\ref{relb})] and correspond to the points
of the Poincar\'e sphere on the $X$-axis (see Eq.\ (\ref{tildevectors}) below). Thus,
the SU(2) matrix $\mathcal{U}$, given by Eq.\ (\ref{u}), represents the connection
between these two frequently employed bases of polarization states (see also Sec.\ 3.1,
below).

Equations (\ref{9}) and (\ref{9c}) constitute a decomposition of a wave as a
superposition of circularly polarized waves, with the components $o^{1}$ and $o^{2}$
being the relative amplitudes of this decomposition. Similarly, $\tilde{o}^{1}$ and
$\tilde{o}^{2}$ are the relative amplitudes of the decomposition of the wave as a
superposition of two linearly polarized waves. (In fact, {\em any}\/ pair of different
points of the Poincar\'e sphere represent a basis; the pairs of points diametrically
opposite are the orthogonal bases \cite{Pa,GF}.)

According to Eq.\ (\ref{relation}), the vectors $R_{i}$ and $M_{i}$ are given in terms
of the spinor $\tilde{o}$ by [see Eqs.\ (\ref{vectors})]
\begin{equation}
R_{i} = \tilde{o}^{\dag} \mathcal{U} \sigma_{i} \mathcal{U}^{-1} \tilde{o}, \qquad
M_{i} = \tilde{o}^{\rm t} \varepsilon \mathcal{U} \sigma_{i} \mathcal{U}^{-1}
\tilde{o}, \label{tildevectors}
\end{equation}
where we have made use of the relation $(\mathcal{U}^{-1})^{\rm t} \varepsilon =
\varepsilon \mathcal{U}$, which applies to unimodular $2 \times 2$ matrices. Equations
(\ref{tildevectors}) are of the same form as Eqs.\ (\ref{vectors}), with $o$ replaced
by $\tilde{o}$ and $\sigma_{i}$ replaced by $\mathcal{U} \sigma_{i} \mathcal{U}^{-1}$.
As shown in Eqs.\ (\ref{paulim}), the matrices $\mathcal{U} \sigma_{i}
\mathcal{U}^{-1}$ are a cyclic permutation of the Pauli matrices (which explains the
definition of the Pauli matrices adopted, without justification, in Ref.\ 11, Appendix
B).

Thus, apart from the factor ${\rm e}^{{\rm i} (kz - \omega t)}$, the components of the
Jones vector (\ref{jones}) are the components of a constant SU(2) spinor (that is,
independent of $t$ and $z$), $\tilde{o}$, in a basis that differs from the standard one
[Eq.\ (\ref{relb})]. The unit spinor $\tilde{o}$ allows us to find the vectors $R_{i}$
and ${\rm Re}\, M_{i}$ that represent the polarization state and phase of the wave on
the Poincar\'e sphere [Eqs.\ (\ref{tildevectors}) and (\ref{paulim})] and, since the
inner product of SU(2) spinors is invariant under SU(2) transformations, the inner
product of the spinors corresponding to two waves with the same wavevector determines
if the waves are in phase according to Pancharatnam's definition \cite{Pa,GF} (see also
Ref.\ 14 and the references cited therein).

\section{Geometrical representation of the effect of optical
filters} Since the state of polarization of a monochromatic electromagnetic plane wave
is represented by a point of the Poincar\'e sphere or, up to a phase factor, by a unit
two-component spinor, {\it e.g.}, $o$ or $\tilde{o}$, the effect of an optical filter
on the polarization of a wave passing through the filter corresponds to some
transformation of the Poincar\'e sphere into itself or to some spinor transformation
(see also Refs.\ 11 and 9).

In this section, following Ref.\ 11, we consider some simple examples of optical
filters, finding their representation on the spinor space and on the Poincar\'e sphere.

\subsection{Phase shifters}
If an optical filter produces a phase shift $\delta_{1}$ for the $x$-component of the
electric field and a, possibly different, phase shift $\delta_{2}$ for the
$y$-component, the electric field (\ref{4}) is replaced by
\begin{eqnarray}
{\bf E} & = & A \left\{ \left[ \cos {\textstyle \frac{1}{2}} \theta \, \cos (\omega t -
kz + {\textstyle \frac{1}{2}} \chi + {\textstyle \frac{1}{2}} \phi + \delta_{1}) + \sin
{\textstyle \frac{1}{2}} \theta \, \cos (\omega t - kz + {\textstyle \frac{1}{2}} \chi
- {\textstyle \frac{1}{2}} \phi + \delta_{1}) \right] \hat{x} \right.
\nonumber \\
& & \mbox{} + \left. \left[ \cos {\textstyle \frac{1}{2}} \theta \, \sin (\omega t - kz
+ {\textstyle \frac{1}{2}} \chi + {\textstyle \frac{1}{2}} \phi + \delta_{2}) - \sin
{\textstyle \frac{1}{2}} \theta \, \sin (\omega t - kz + {\textstyle \frac{1}{2}} \chi
- {\textstyle \frac{1}{2}} \phi + \delta_{2}) \right] \hat{y}
\right\}. \nonumber \\
& & \label{4shift}
\end{eqnarray}
This expression is equivalent to
\begin{eqnarray*}
E_{x} + {\rm i} E_{y} & = & A \left\{ {\rm e}^{{\rm i} (\omega t - kz + (\delta_{1} +
\delta_{2})/2)} [\cos {\textstyle \frac{1}{2}} \delta \cos {\textstyle \frac{1}{2}}
\theta \, {\rm e}^{{\rm i} (\chi + \phi)/2} - {\rm i} \sin {\textstyle \frac{1}{2}}
\delta \sin {\textstyle \frac{1}{2}} \theta \, {\rm e}^{{\rm i} (\chi -
\phi)/2}] \right. \\
& & \left. \mbox{} + {\rm e}^{- {\rm i} (\omega t - kz + (\delta_{1} + \delta_{2})/2)}
[\cos {\textstyle \frac{1}{2}} \delta \sin {\textstyle \frac{1}{2}} \theta \, {\rm
e}^{- {\rm i} (\chi - \phi)/2} + {\rm i} \sin {\textstyle \frac{1}{2}} \delta \cos
{\textstyle \frac{1}{2}} \theta \, {\rm e}^{- {\rm i} (\chi + \phi)/2}] \right\},
\end{eqnarray*}
which is duly of the form (\ref{9}), with the two-component spinor $o$ replaced by
\begin{equation}
\left( \begin{array}{c} o'{}^{1} \\ o'{}^{2} \end{array} \right) = {\rm e}^{- {\rm i}
(\delta_{1} + \delta_{2})/2} \left(
\begin{array}{cc} \cos \frac{1}{2} \delta & {\rm i} \sin \frac{1}{2}
\delta \\ {\rm i} \sin \frac{1}{2} \delta & \cos \frac{1}{2} \delta
\end{array} \right) \left( \begin{array}{c} o^{1} \\ o^{2}
\end{array} \right), \label{mshift}
\end{equation}
where $\delta \equiv \delta_{2} - \delta_{1}$.

Apart from the overall phase factor ${\rm e}^{- {\rm i} (\delta_{1} + \delta_{2})/2}$,
the transformation (\ref{mshift}) is given by the SU(2) matrix
\begin{equation}
\left( \begin{array}{cc} \cos \frac{1}{2} \delta & {\rm i} \sin \frac{1}{2} \delta \\
{\rm i} \sin \frac{1}{2} \delta & \cos \frac{1}{2} \delta \end{array} \right) = (\cos
{\textstyle \frac{1}{2}} \delta) I + {\rm i} (\sin {\textstyle \frac{1}{2}} \delta) \,
\sigma_{1} = \exp ({\rm i} {\textstyle \frac{1}{2}} \delta \, \sigma_{1}),
\label{mshiftb}
\end{equation}
where $I$ is the $2 \times 2$ identity matrix, which corresponds to a rotation on the
Poincar\'e sphere through an angle $- \delta$ about the $X$-axis.

There exist two diametrically opposite points of the Poincar\'e sphere that are
invariant under this rotation (the points on the intersection of the Poincar\'e sphere
and the $X$-axis), which, therefore, correspond to polarization states that are not
affected by this filter. These two polarization states are linearly polarized waves
with the electric field along the $x$-axis or the $y$-axis [the states (\ref{linb})],
as one would expect. (Note that, owing to the definition of the angle $\phi$ given in
Sec.\ 2, a rotation of the coordinate axes in the $xy$-plane through an angle $\alpha$
produces the substitution of $\phi/2$ by $(\phi/2) - \alpha$, which corresponds to the
action of the matrix
\begin{equation}
\left( \begin{array}{cc} {\rm e}^{{\rm i} \alpha} & 0 \\ 0 & {\rm e}^{- {\rm i} \alpha}
\end{array} \right) = \exp ({\rm i} \alpha \, \sigma_{3}) \label{sn}
\end{equation}
on the spinor $o$. This SU(2) matrix, in turn, corresponds to a rotation on the
Poincar\'e sphere through an angle $- 2 \alpha$ about the $Z$-axis. Thus, a rotation by
$90^{\circ}$ in the $xy$-plane, which transforms a linear polarization along the
$x$-axis into a linear polarization along the $y$-axis, corresponds to a rotation by
$180^{\circ}$ in the Poincar\'e sphere.)

According to Eqs.\ (\ref{paulim}), with respect to the basis (\ref{linb}), formed by
linearly polarized states, the spinor transformation (\ref{mshift}) is given by the
unitary matrix
\begin{equation}
{\rm e}^{- {\rm i} (\delta_{1} + \delta_{2})/2} \exp ({\rm i} {\textstyle \frac{1}{2}}
\delta \, \sigma_{3}) = {\rm e}^{- {\rm i} (\delta_{1} + \delta_{2})/2} \left(
\begin{array}{cc} {\rm e}^{{\rm i} \delta/2} & 0 \\ 0 & {\rm e}^{- {\rm i} \delta/2}
\end{array} \right) = \left( \begin{array}{cc} {\rm e}^{- {\rm i} \delta_{1}} & 0 \\ 0
& {\rm e}^{- {\rm i} \delta_{2}} \end{array} \right), \label{mshiftbl}
\end{equation}
as one would expect, owing to the definition of $\delta_{1}$ and $\delta_{2}$.

In order to reduce the possible confusions coming from the simultaneous use of two
different bases, it is convenient to make use of Dirac's notation, denoting by $| +
\rangle$ and $| - \rangle$ the states with circular polarization (\ref{circb}),
respectively. Then,
\begin{eqnarray}
| x \rangle & \equiv & \frac{1}{\sqrt{2}} {\rm e}^{- {\rm i} \pi /4} \; | + \rangle +
\frac{1}{\sqrt{2}} {\rm e}^{- {\rm i} \pi /4} \;
| - \rangle, \nonumber \\
| y \rangle & \equiv & - \frac{1}{\sqrt{2}} {\rm e}^{{\rm i} \pi /4} \; | + \rangle +
\frac{1}{\sqrt{2}} {\rm e}^{{\rm i} \pi /4} \; | - \rangle, \label{ketbases}
\end{eqnarray}
correspond to states with linear polarization (the states (\ref{linb}), which are
essentially the states $| \textsf{v} \rangle$ and $| \textsf{h} \rangle$ with vertical
and horizontal polarization employed in Ref.\ 5). (See Eq.\ (\ref{u}).) In this manner,
the SU(2) transformation (\ref{mshiftb}) is expressed as
\[
(\cos {\textstyle \frac{1}{2}} \delta) I + {\rm i} (\sin {\textstyle \frac{1}{2}}
\delta) \big( | + \rangle \langle - | \; + \; | - \rangle \langle + | \big),
\]
which, by virtue of Eqs.\ (\ref{ketbases}), amounts to
\begin{equation}
|x \rangle {\rm e}^{{\rm i} \delta/2} \langle x| + |y \rangle {\rm e}^{-{\rm i}
\delta/2} \langle y| \label{ddiag}
\end{equation}
and corresponds to the diagonal matrix ${\rm diag}\, ({\rm e}^{{\rm i} \delta/2}, {\rm
e}^{-{\rm i} \delta/2})$ appearing in Eq.\ (\ref{mshiftbl}).

The effect represented by the SU(2) transformation (\ref{ddiag}) comes from the
anisotropy of the medium, which produces different effects on the linearly polarized
waves with electric field along the $x$-axis or the $y$-axis. In an analogous manner, a
{\em gyrotropic}\/ medium (see, {\it e.g.}, Ref.\ 15) produces different effects on the
waves with right or left circular polarization; therefore, the effect of a gyrotropic
medium is represented by
\[
|+ \rangle {\rm e}^{-{\rm i} \delta_{1}} \langle +| + |- \rangle {\rm e}^{-{\rm i}
\delta_{2}} \langle -| = {\rm e}^{- {\rm i} (\delta_{1} + \delta_{2})/2} \big( |+
\rangle {\rm e}^{{\rm i} \delta/2} \langle +| + |- \rangle {\rm e}^{-{\rm i} \delta/2}
\langle -| \big),
\]
where $\delta \equiv \delta_{2} - \delta_{1}$, or by the unitary matrix ${\rm e}^{-
{\rm i} (\delta_{1} + \delta_{2})/2} \exp ({\rm i} {\textstyle \frac{1}{2}} \delta \,
\sigma_{3})$, which corresponds to a rotation on the Poincar\'e sphere through an angle
$- \delta$ about the $Z$-axis.

Hence, with respect to the basis (\ref{linb}), formed by states with linear
polarization, making use of Eqs.\ (\ref{paulim}) or (\ref{ketbases}), the effect of a
gyrotropic medium will be represented by a matrix of the form
\begin{equation}
{\rm e}^{- {\rm i} (\delta_{1} + \delta_{2})/2} \exp ({\rm i} {\textstyle \frac{1}{2}}
\delta \, \sigma_{2}) = {\rm e}^{- {\rm i} (\delta_{1} + \delta_{2})/2} \left(
\begin{array}{cr} \cos \frac{1}{2} \delta & - \sin \frac{1}{2} \delta \\ \sin
\frac{1}{2} \delta & \cos \frac{1}{2} \delta \end{array} \right). \label{gyrl}
\end{equation}

A quarter-wave plate \cite{SM} is a phase shifter corresponding to a rotation on the
Poincar\'e sphere through $\pi/2$ about an axis on the $XY$-plane. Hence, with respect
to the basis $\{ |+\rangle, |-\rangle \}$, it is represented by the SU(2) matrix
\[
(\cos \pi/4) I - {\rm i} (\sin \pi/4) [(\cos 2 \theta) \, \sigma_{1} + (\sin 2 \theta)
\, \sigma_{2}] = \frac{1}{\sqrt{2}} [I - {\rm i} (\cos 2 \theta) \, \sigma_{1} - {\rm
i} (\sin 2 \theta) \, \sigma_{2}],
\]
where $\theta$ is the angle between the axis of the plate and the $x$-axis [see the
discussion after Eq.\ (\ref{sn})], and, according to Eqs.\ (\ref{paulim}), with respect
to the basis $\{ |x \rangle, |y \rangle \}$, it is represented by
\[
\frac{1}{\sqrt{2}} [I - {\rm i} (\cos 2 \theta) \, \sigma_{3} - {\rm i} (\sin 2 \theta)
\, \sigma_{1}].
\]
A half-wave plate is a phase shifter corresponding to a rotation on the Poincar\'e
sphere through $\pi$ about an axis on the $XY$-plane and, therefore, is represented by
the square of the matrix corresponding to a quarter-wave plate.

\subsection{Attenuators}
In the case of an optical filter that produces an attenuation given by a factor ${\rm
e}^{- \eta_{1}}$ for the $x$-component of the electric field and an attenuation given
by ${\rm e}^{- \eta_{2}}$ for the $y$-component, the electric field (\ref{4}) is
replaced by
\begin{eqnarray}
{\bf E} & = & A \left\{ {\rm e}^{- \eta_{1}} \left[ \cos {\textstyle \frac{1}{2}}
\theta \, \cos (\omega t - kz + {\textstyle \frac{1}{2}} \chi + {\textstyle
\frac{1}{2}} \phi) + \sin {\textstyle \frac{1}{2}} \theta \, \cos (\omega t - kz +
{\textstyle \frac{1}{2}} \chi - {\textstyle \frac{1}{2}} \phi) \right] \hat{x}
\right. \nonumber \\
& & \mbox{} + \left. {\rm e}^{- \eta_{2}} \left[ \cos {\textstyle \frac{1}{2}} \theta
\, \sin (\omega t - kz + {\textstyle \frac{1}{2}} \chi + {\textstyle \frac{1}{2}} \phi)
- \sin {\textstyle \frac{1}{2}} \theta \, \sin (\omega t - kz + {\textstyle
\frac{1}{2}} \chi - {\textstyle \frac{1}{2}} \phi) \right] \hat{y} \right\}.
\label{4atte}
\end{eqnarray}
This expression is equivalent to
\begin{eqnarray*}
E_{x} + {\rm i} E_{y} & = & A {\rm e}^{-(\eta_{1} + \eta_{2})/2} \left\{ {\rm e}^{{\rm
i} (\omega t - kz)} [\cosh {\textstyle \frac{1}{2}} \eta \cos {\textstyle \frac{1}{2}}
\theta \, {\rm e}^{{\rm i} (\chi + \phi)/2} + \sinh {\textstyle \frac{1}{2}} \eta \sin
{\textstyle \frac{1}{2}} \theta \, {\rm e}^{{\rm i} (\chi -
\phi)/2}] \right. \\
& & \left. \mbox{} + {\rm e}^{- {\rm i} (\omega t - kz)} [\cosh {\textstyle
\frac{1}{2}} \eta \sin {\textstyle \frac{1}{2}} \theta \, {\rm e}^{- {\rm i} (\chi -
\phi)/2} + \sinh {\textstyle \frac{1}{2}} \eta \cos {\textstyle \frac{1}{2}} \theta \,
{\rm e}^{- {\rm i} (\chi + \phi)/2}] \right\},
\end{eqnarray*}
which is of the form (\ref{9}), with the two-component spinor $o$ replaced by
\begin{equation}
\left( \begin{array}{c} o'{}^{1} \\ o'{}^{2} \end{array} \right) = {\rm e}^{- (\eta_{1}
+ \eta_{2})/2} \left( \begin{array}{cc} \cosh \frac{1}{2} \eta & \sinh \frac{1}{2} \eta
\\ \sinh \frac{1}{2} \eta & \cosh \frac{1}{2} \eta \end{array} \right) \left(
\begin{array}{c} o^{1} \\ o^{2} \end{array} \right) = {\rm e}^{- (\eta_{1} +
\eta_{2})/2} \exp ({\textstyle \frac{1}{2}} \eta \, \sigma_{1}) \left(
\begin{array}{c} o^{1} \\ o^{2} \end{array} \right), \label{matte}
\end{equation}
where $\eta \equiv \eta_{2} - \eta_{1}$. The $2 \times 2$ matrix appearing in Eq.\
(\ref{matte}) is unimodular, but does not belong to SU(2) and, therefore, it does not
correspond to a rotation on the Poincar\'e sphere. Rather, it corresponds to a
conformal transformation of the sphere (see, {\it e.g.}, Ref.\ 16). In any case, the
effect of the attenuator on the polarization state of a wave is represented by a
transformation on the points of the Poincar\'e sphere.

Clearly, if there is an attenuation given by a factor ${\rm e}^{- \eta_{1}}$ for the
$x$-component of the electric field and an attenuation given by ${\rm e}^{- \eta_{2}}$
for the $y$-component, the column matrix (\ref{jones}) is replaced by
\begin{equation}
\left( \begin{array}{c} E'_{x} \\ E'_{y} \end{array} \right) = {\rm e}^{-(\eta_{1} +
\eta_{2})/2} \left( \begin{array}{cc} {\rm e}^{\eta/2} & 0 \\ 0 & {\rm e}^{- \eta/2}
\end{array} \right) \left(
\begin{array}{c} E_{x} \\ E_{y} \end{array} \right),
\end{equation}
and the non-unitary, unimodular matrix appearing in this last equation, which can be
expressed as $\exp (\frac{1}{2} \eta \, \sigma_{3})$, is exactly what we should expect
taking into account Eqs.\ (\ref{matte}) and (\ref{paulim}).

\section{Partially polarized beams}
As is often remarked, by contrast with the Jones vector, the Stokes parameters can also
be used to deal with partially polarized beams. In this section we show that the
two-component spinor formalism can be easily adapted to handle partially polarized
light, and, as we shall see, the resulting description is equivalent to that given by
the coherency matrix ({\it cf.}\ Ref.\ 11, Appendix B).

The Stokes parameters allow us to distinguish a completely polarized beam from a
partially polarized beam. Letting
\begin{equation}
S \equiv s_{0}{}^{2} - s_{1}{}^{2} - s_{2}{}^{2} - s_{3}{}^{2}, \label{S}
\end{equation}
it turns out that for a completely polarized beam, $S = 0$ [{\it cf.}\ Eq.\
(\ref{stokes})], while for a partially polarized beam, $S > 0$ (see, {\it e.g.}, Ref.\
6, Sec.\ 10.8.3). The four Stokes parameters can be related to a $2 \times 2$ Hermitean
matrix, $C$, by means of
\begin{equation}
s_{\alpha} = {\rm tr} \, (C \sigma_{\alpha}), \qquad (\alpha = 0, 1, 2, 3) \label{C}
\end{equation}
where tr denotes the trace, $\sigma_{0} \equiv I$, and $\sigma_{1}$, $\sigma_{2}$,
$\sigma_{3}$, are the Pauli matrices, as above.

The Hermitean matrix $\rho \equiv C/s_{0}$ has the usual properties of a density matrix
(or density operator) as defined in Quantum Mechanics (see, {\it e.g.}, Ref.\ 17),
namely
\begin{equation}
{\rm tr} \, \rho = 1, \qquad {\rm tr} \, \rho^{2} \leqslant 1. \label{density}
\end{equation}
In fact, Eqs.\ (\ref{C}) (together with the condition $C^{\dag} = C$) are equivalent to
\begin{equation}
C = \frac{1}{2} \left( \begin{array}{cc} s_{0} + s_{3} & s_{1} - {\rm i} s_{2} \\ s_{1}
+ {\rm i} s_{2} & s_{0} - s_{3} \end{array} \right) \label{Cinv}
\end{equation}
that is,
\begin{equation}
C = \frac{1}{2} \sum_{\alpha = 0}^{3} s_{\alpha} \sigma_{\alpha} \label{Cinvi}
\end{equation}
and one readily verifies that ${\rm tr} \, C = s_{0}$, and ${\rm tr} \, C^{2} =
s_{0}{}^{2} - S/2$, which amount to Eqs.\ (\ref{density}), taking into account that $S
\geqslant 0$.

Furthermore, $\det C = S/4$; hence, in the case of a completely polarized wave ($S =
0$), the matrix $C$, having determinant equal to zero, must be of the form $\psi
\psi^{\dag}$, where $\psi$ is some two-component spinor. In fact, writing $C = s_{0} o
o^{\dag}$, where $o$ is a normalized spinor, we recover the (``pure state'') case
considered in Sec.\ 2. Indeed,
\[
{\rm tr} \, (C \sigma_{0}) = s_{0} {\rm tr} \, (o o^{\dag}) = s_{0} o^{\dag} o = s_{0},
\]
and
\[
{\rm tr} \, (C \sigma_{i}) = s_{0} {\rm tr} \, (o o^{\dag} \sigma_{i}) = s_{0} o^{\dag}
\sigma_{i} o = s_{i}, \qquad (i = 1,2,3)
\]
[see Eq.\ (\ref{11})], reproducing Eqs.\ (\ref{C}).

The matrix $C$, being Hermitean, possesses two mutually orthogonal unit eigenspinors
with real eigenvalues. These unit spinors correspond to two diametrically opposite
points of the Poincar\'e sphere [see Ref.\ 1, Eq.\ (18)]. Since $C$ is a $2 \times 2$
matrix, its two eigenvalues coincide only when $C$ is a multiple of the identity matrix
and, only in this case, which corresponds to ``unpolarized'' light ($s_{1} = s_{2} =
s_{3} = 0$), the direction of the eigenspinors of $C$ is not uniquely defined. In all
cases, the unit eigenspinors of $C$ are defined up to a phase factor, hence, there are
no uniquely defined tangent vectors to the Poincar\'e sphere at these points, analogous
to the vector ${\rm Re}\, {\bf M}$ defined in Sec.\ 2.

Thus, in the case of a partially polarized beam (a ``mixed state''), the polarization
state defines two diametrically opposite points of the Poincar\'e sphere (except in the
case of unpolarized light). However, these two points (which correspond to the
eigenspinors of $C$) do not fully specify the matrix $C$, since the eigenvalues need to
be known. According to the discussion in Sec.\ 2, the vectors $\pm (s_{1}, s_{2},
s_{3})$ point along the directions of the points of the Poincar\'e sphere representing
the partially polarized beam.

When the beam is completely polarized, $C$ is of the form $C = s_{0} o o^{\dag}$; the
unit spinor $o$ is an eigenspinor of $C$ ($C o = s_{0} o o^{\dag} o = s_{0} o$) and any
spinor orthogonal to $o$ ({\it e.g.}, the mate of $o$ \cite{GF}) is also an eigenspinor
of $C$ (with eigenvalue equal to zero).

As with any matrix, the form and properties of $C$ depend on the basis employed.
Fortunately, making use of Eq.\ (\ref{Cinvi}), which gives $C$ in terms of the Pauli
matrices, and Eqs.\ (\ref{paulim}), we can obtain at once the expression of $C$ in the
basis formed by the unit spinors (\ref{linb}); the resulting expression is
\begin{equation}
\widetilde{C} = \frac{1}{2} (s_{0} I + s_{1} \sigma_{3} + s_{2} \sigma_{1} + s_{3}
\sigma_{2}) = \frac{1}{2} \left(
\begin{array}{cc} s_{0} + s_{1} & s_{2} - {\rm i} s_{3} \\ s_{2}
+ {\rm i} s_{3} & s_{0} - s_{1} \end{array} \right). \label{ctil}
\end{equation}
Taking into account the relationship between the Stokes parameters and the elements of
the coherency matrix, $J_{ij}$ (see, {\it e.g.}, Ref.\ 6, Sec.\ 10.8.3), we have
\begin{equation}
\widetilde{C} = \left( \begin{array}{cc} J_{xx} & J_{yx}
\\ J_{xy} & J_{yy} \end{array} \right).
\label{cohe}
\end{equation}

When a partially polarized beam passes through a phase shifter, the matrix $C$,
corresponding to the initial beam, is replaced by $QCQ^{\dag}$, where $Q$ is the SU(2)
matrix representing the effect of the filter on the state of polarization ($\exp ({\rm
i} \frac{1}{2} \delta \, \sigma_{1})$ or $\exp ({\rm i} \frac{1}{2} \delta \,
\sigma_{3})$ in the cases considered in Sec.\ 3.1; note that the factors ${\rm e}^{-
{\rm i} (\delta_{1} + \delta_{2})/2}$ appearing in Eqs.\ (\ref{mshift}) and
(\ref{gyrl}) are not present in $QCQ^{\dag}$ because they have unit modulus). The
eigenspinors of $QCQ^{\dag}$ are the images under $Q$ of those of $C$; therefore, the
diametrically opposite points on the Poincar\'e sphere defined by $QCQ^{\dag}$ are
obtained from those defined by $C$ by means of the rotation corresponding to $Q$ (see
also Refs.\ 18 and 19).

Similarly, when a partially polarized beam passes through an attenuator, the initial
matrix $C$ is transformed into
\[
{\rm e}^{- (\eta_{1} + \eta_{2})/2} \exp ({\textstyle \frac{1}{2}} \eta \, \sigma_{1})
\, C \,[{\rm e}^{- (\eta_{1} + \eta_{2})/2} \exp ({\textstyle \frac{1}{2}} \eta \,
\sigma_{1})]^{\dag} = {\rm e}^{- (\eta_{1} + \eta_{2})} \exp ({\textstyle \frac{1}{2}}
\eta \, \sigma_{1}) \, C \, \exp ({\textstyle \frac{1}{2}} \eta \, \sigma_{1})
\]
[see Eq.\ (\ref{matte})], which is of the form (\ref{Cinvi}), with $(s_{0}, s_{1},
s_{2}, s_{3})$ replaced by
\begin{eqnarray}
{\rm e}^{- (\eta_{1} + \eta_{2})} (s_{0} \cosh \eta + s_{1} \sinh \eta, s_{1} \cosh
\eta + s_{0} \sinh \eta, s_{2}, s_{3}).
\end{eqnarray}
Thus, apart from the overall factor ${\rm e}^{- (\eta_{1} + \eta_{2})}$, the effect of
an attenuator on the Stokes parameters has the form of a Lorentz boost in the
$x$-direction (see also Refs.\ 10 and 11).

\section{Conclusions}
We have shown that the several objects and formalisms employed in the study of the
polarization of electromagnetic waves are deeply related, despite their apparent
differences. In particular, the identification of the Jones vector with a SU(2) spinor,
allows us to represent the Jones vector by a tangent vector to the Poincar\'e sphere,
in terms of which, among other things, the Pancharatnam phase can be visualized.

\section*{Acknowledgments}
One of the authors (I.R.G.) would like to thank the Departamento de F\'isica
Matem\'atica of the Instituto de Ciencias, Universidad Aut\'onoma de Puebla for its
kind hospitality, the Sistema Nacional de Investigadores (M\'exico) and the Cuerpo
acad\'emico de part\'iculas, campos y relatividad general (U.A.P.) for financial
support. The authors would also like to thank the referee for useful comments and for
pointing out Refs.\ 12, 18, and 19 to them.





\begin{thebibliography}{99}
\bibitem{GF} G. F. Torres del Castillo, J.\ Phys.\ A: Math.\ Theor.\
{\bf 41} (2008) 115302.
\bibitem{WP} W. T. Payne, Am.\ J.\ Phys.\ {\bf 20} (1952) 253.
\bibitem{Pa} S. Pancharatnam, Proc.\ Ind.\ Acad.\ Sci.\ A {\bf 44} (1956)
247. Reprinted in {\it Geometric Phases in Physics}, edited by A. Shapere and F.
Wilczek (World Scientific, Singapore, 1989).
\bibitem{Be} M. V. Berry, J.\ Mod.\ Optics {\bf 34} (1987) 1401. Reprinted in {\it Geometric Phases in Physics}, edited
by A. Shapere and F. Wilczek (World Scientific, Singapore, 1989).
\bibitem{En} B.-G. Englert, C. Kurtsiefer, and H. Weinfurter,
Phys.\ Rev.\ A {\bf 63} (2001) 032303.
\bibitem{BW} M. Born and E. Wolf, {\it Principles of Optics}, 6th
ed.\ (Cambridge University Press, Cambridge, 1997).
\bibitem{Gu} R. D. Guenther, {\it Modern Optics} (Wiley, New York,
1990), Chap.\ 2.
\bibitem{KL} D. S. Kliger and J. W. Lewis, {\it Polarized Light in
Optics and Spectroscopy} (Academic Press, New York, 1990).
\bibitem{SM} R. Simon and N. Mukunda, Phys.\ Lett.\ A {\bf 143}
(1990) 165.
\bibitem{HK} D. Han, Y. S. Kim, and M. E. Noz, Phys.\ Rev.\ E {\bf 56} (1997)
6065.
\bibitem{Jo} D. Han, Y. S. Kim, and M. E. Noz, J.\ Opt.\ Soc.\ Am.\ A
{\bf 14} (1997) 2290.
\bibitem{Tu} T.\ Tudor, J.\ Phys.\ A: Math.\ Theor.\ {\bf 41} (2008) 415303.
\bibitem{3D} G. F. Torres del Castillo, {\it 3-D Spinors, Spin-weighted
Functions and their Applications} (Birkh\"auser, Boston, 2003).
\bibitem{Lo} J. C. Loredo, O. Ort\'iz, R. Weing\"artner, and F. De
Zela, Phys.\ Rev.\ A {\bf 80} (2009) 012113.
\bibitem{LL} L. D. Landau, E. M. Lifshitz, and L. P. Pitaevski\u{\i},
{\it Electrodynamics of Continuous Media}, 2nd ed.\ (Pergamon Press, Oxford, 1984),
\S101.
\bibitem{PR} R. Penrose and W. Rindler, {\it Spinors and
space-time}, Vol.\ 1 (Cambridge University Press, Cambridge, 1984), Sec.\ 1.3.
\bibitem{Go} K. Gottfried and T.-M. Yan, {\it Quantum Mechanics:
Fundamentals}, 2nd ed.\ (Springer-Verlag, New York, 2003), Sec.\ 2.2.
\bibitem{TM} T. Tudor and V. Manea, J.\ Opt.\ Soc.\ Am.\ B {\bf 28} (2011) 596.
\bibitem{Gi} J.J.\ Gil, Eur.\ Phys.\ J.\ Appl.\ Phys.\ {\bf 40} (2007) 1.

\end{thebibliography}
\end{document}